\documentclass[conference]{IEEEtran}
\IEEEoverridecommandlockouts
\usepackage{cite}
\usepackage{amsmath,amssymb,amsfonts}
\usepackage{algorithmic}
\usepackage{graphicx}
\usepackage{textcomp}
\usepackage{xcolor}
\def\BibTeX{{\rm B\kern-.05em{\sc i\kern-.025em b}\kern-.08em
    T\kern-.1667em\lower.7ex\hbox{E}\kern-.125emX}}
\begin{document}
\raggedbottom
\title{An interdisciplinary overview of developmental indices and behavioral measures of the minimal self
\thanks{This work was funded by the Deutsche Forschungsgemeinschaft (DFG, German Research Foundation) - 402790442 ("Prerequisites for the Development of an Artificial Self") within the SPP "The Active Self" (SPP 2134). The work of GS has partially received funding from the European Union’s Horizon 2020 research and innovation programme under the Marie Sklodowska-Curie grant agreement No. 838861 (Predictive Robots).}
}

\author{
\IEEEauthorblockN{Yasmin Kim Georgie}
\IEEEauthorblockA{\textit{Adaptive Systems Group} \\
\textit{Humboldt-Universit\"{a}t zu Berlin}\\
Berlin, Germany \\
yasmin.kim.georgie@informatik.hu-berlin.de}
\and
\IEEEauthorblockN{Guido Schillaci}
\IEEEauthorblockA{\textit{The BioRobotics Institute}\\ \textit{Scuola Superiore Sant'Anna}, Pisa, Italy\\
and \textit{Adaptive Systems Group} \\
\textit{Humboldt-Universit\"{a}t zu Berlin}\\
guido.schillaci@santannapisa.it}
\and
\IEEEauthorblockN{Verena Vanessa Hafner}
\IEEEauthorblockA{\textit{Adaptive Systems Group} \\
\textit{Humboldt-Universit\"{a}t zu Berlin}\\
Berlin, Germany \\
hafner@informatik.hu-berlin.de}
}

\maketitle

\begin{abstract}

In this review paper we discuss the development of the minimal self in humans, the behavioural measures indicating the presence of different aspects of the minimal self, namely, body ownership and sense of agency, and also discuss robotics research investigating and developing these concepts in artificial agents. We investigate possible avenues for expanding the research in robotics to further explore the development of an artificial minimal self.  
\end{abstract}

\begin{IEEEkeywords}
Models of self and agency; Sensorimotor development; Machine Learning methods for robot development
\end{IEEEkeywords}

\section{Introduction}
For centuries, philosophers and scholars from different disciplines have been debating the nature of subjective experience and self-consciousness.
A recent account brings back a phenomenological view on this debate, researching self-consciousness in its minimal form, that is studying subjective experiences in their immediate and first-personal "mineness". According to Gallagher, this so-called \textit{minimal self} refers to the "consciousness of oneself as an immediate subject of experience, unextended in time" \cite{gallagher2000philosophical}. Aspects of this minimal self involve the sense of \textit{agency}---the sense of the self as the one causing or generating an action, and the sense of \textit{ownership}---the sense of the self as the one subjected to an experience \cite{gallagher2000philosophical}. 

This view is distinct from more elaborated aspects of the self, such as the reflexive self and the narrative self \cite{martin2014temporal}. The minimal self is closer to a minimalist level of subjective experience, where the focus is more on the contribution of the here and now, bodily experience in its construction \cite{ferri2011motor}. Other low-level notions of the self have been proposed in the literature, such as the proto-self and the immunological self \cite{damasio2003mental}. Hereby, we do not enter the debate about which of them constitutes the lowest level of consciousness, as researchers have not yet converged to an agreement.
We commit to the notion of the minimal self, as this aspect of consciousness is perhaps the most easily accessible in terms of experimental exploration and quantification, and is in fact receiving greater attention from disciplines such as neuroscience, behavioural and cognitive sciences, and developmental psychology \cite{engel2001temporal, crick1990towards}. 

Developmental psychologists consider the emergence of a sense of the self as a key step in cognitive development. By the second year of life, thus few months after having acquired basic linguistic skills, toddlers are capable of using self-referential language such as \textit{I}, \textit{me}, \textit{my}---suggesting that the acquisition of a \textit{self-concept} has started earlier. The minimal self is pre-linguistic and non-conceptual, and is suggested to unfold already during early developmental stages \cite{rochat2003five}. 

This paper presents an interdisciplinary overview of developmental indices and behavioural measures of the minimal self. 
The minimal self is argued to include two main aspects---a sense of agency and a sense of body ownership---which are thought to be dependent on an internal body representation maintained by our brain. This manuscript thus starts with discussing the development of body representations as a necessary condition for the emergence of the senses of body ownership and of agency (section \ref{sec:development}). Behavioural paradigms and measures indicating the presence of different aspects of the minimal self (section \ref{sec:measures}) are analysed. In particular, we survey studies on self-touch, intentional binding and sensory attenuation, and the rubber hand illusion. 

Alongside the survey on the developmental aspects of the minimal self and the aforementioned behavioural paradigms and measures, the main goal of this work is to discuss the most prominent related studies in robotics. In fact, there is a growing interest in the developmental robotics community in implementing processes capable of enabling the experience of the self in artificial agents. Self-awareness may clearly improve adaptivity and autonomy in robots and, as a result, reduce human intervention in their programming. Using robots as test-beds for studying the minimal self may also shed light on the cognitive mechanisms underling subjective experience.
Nonetheless, the investigation on the artificial self is still young and fragmented. This work contributes with identifying current knowledge gaps and limitations in robotics studies and with suggesting research directions for the implementation of behavioural paradigms and measures for the artificial self.








\section{The development of body representations}
\label{sec:development}
In order to effectively interact with the environment, an embodied agent must form and maintain an internal representation of its own body situated within the environment. The term \textit{representation} may sound controversial. In fact, scholars usually take either a representationalist or a sensorimotor position in the philosophical study of bodily awareness (see \cite{sep-bodily-awareness}, section 2, for a review), which may lead to the definition of body \textit{image}---the mental representation of the body, constituted by a combination of sensorimotor experience and social and psychological concepts about it, and body \textit{schema}---the integrated, neural organisation of multimodal stimuli coming from the different parts of the body, which is essential for movements. Here, we commit to the latter interpretation of the term body representation, that is a mapping of the body in its various modalities (tactile, proprioceptive, visual, motor, etc.), which is operated in a nonconscious way \cite{gallagher1986body}.
The neural foundation for these representations are the so-called cortical "homunculi" in the primary sensory (S1) and motor (M1) cortices. These are neurological representations of the different anatomical divisions of the body, mapped onto brain areas charged with sensory and motor processing along S1 and M1, respectively.
These specialized areas are organised in a somatotopic map where adjacent body parts are represented closely together. The extent of cortex dedicated to a body region is not proportional to its size in the body, but rather to the density of innervation in that specific part (e.g. the mouth and palms).
The establishment of the somatotopic organisation in S1 and M1 is driven by genetic factors that are later elaborated through changes in connectivity driven by embodied interactions both before and after birth \cite{dall2018somatotopic}.  

Body representations dynamically integrate information from different sensory modalities: tactile, proprioceptive, vestibular, motor, and visual.
Studies suggest that the first sense to emerge in the foetus is the somatosensory sense\cite{bradley1975fetal}, where foetuses are in a state of constantly being touched by their environment. In addition, they engage in self-touch in the womb: often touching their mouth and feet---body parts that are highly innervated and therefore most sensitive to touch---and later on, other parts of the body. The early inclination for movements and self-touch in parts of the body that are more sensitive, suggests that the foetus shows a preference towards movements that induce more informative sensations \cite{fagard2018fetal}

From as early as 19 weeks, foetuses can anticipate hand-to-mouth touch, in opening their mouth prior to contact \cite{myowa2006human, reissland2014development}, indicating the early presence of a sort of sensorimotor mapping and inference.
From 22 weeks, movements seem to show a form of intentional movement, as they become more direct dependening on the action goal \cite{zoia2007evidence}. 
Evidence from neural development studies suggests that even before birth, the prenatal brain should be able to perceive information arising from the body, while higher level (multimodal) representations are possibly formed during the first year after birth, in accordance with the development of association areas \cite{hoffmann2017role}. 

At around 2 months of age, the dominant control of behaviours transitions from subcortical to higher order cortical systems \cite{mcgraw1943neuromuscular}: PET studies show dominant metabolic activity in subcortical regions and the sensorimotor cortex in infants under 5 weeks after birth, and by 3 months, an increase in metabolic activity in the parietal, temporal, and dorsolateral occipital cortices \cite{chugani1994development}. 
Hand-mouth coordination continues to develop after birth, and from birth to 6 months, infants display self-touch in a progressive manner throughout their body, from frequently touching rostral parts such as the head and trunk, to more caudal parts of the body such as the hips, legs, and feet later on \cite{thomas2015independent}.
The development of goal-directed reaching considerably speeds up at about 5 months of age.
Reaching to the own body is thought to be the product of interactions between multiple subsystems. The body representation is constructed through reaching to the body because, in this, the body that is used to act upon the environment becomes the target itself, and therefore needs to be modeled. In certain cases, the self-touch process seems to bypass vision, as when the target is the face, relying only on somatosensory information. 

\textit{In summary}, genetically predetermined cortical maps---the "homunculi" in the primary sensory (S1) and motor (M1) cortices---facilitate the formation of body representations through cortical learning of sensorimotor contingencies---i.e. the statistical connections between sensory and motor information, and sensorimotor integration---integrating this information into common percepts.
This learning of sensorimotor contingencies both drives and is rendered through interactions between brain-body-environment. Specifically, interlinked with the neural ontogenetic process (brain maturation, brain-body interaction), self-exploration (body babbling), self-touch, and goal-directed reaching are considered the necessary behavioural conditions that facilitate and reflect this process. 
Importantly, this process is thought to be driven and progressively refined by the reduction of prediction errors between predicted sensory outcomes and motor actions, such that the agent learns not only to predict the outcomes of its actions on the environment, but also to predict the (sensory) outcomes of its (motor) actions on its own body \cite{blakemore1999spatio,friston2005theory}. 


Rochat \cite{rochat1998self} describes the idea that infants' self-exploration, and interactions with the environment, give rise to the sense of body ownership through the "ability to detect intermodal invariants and regularities in their sensorimotor experience, which specify themselves as separate entities agent in the environment." Therefore, self-exploration (body babbling), self-touch, and goal-directed reaching are necessary conditions for the development of motor control and the emergence of body ownership and sense of agency.

\section{Behavioural paradigms and measures of the self}
\label{sec:measures}
A challenging task in the study of the minimal self is to experimentally  quantify the attribution of subjective experience. A number of behavioural paradigms and attempts for objective measures have been proposed. This section reviews the most prominent ones. 
In particular, we analyse studies on self-touch, intentional binding and sensory attenuation, and the rubber hand illusion.
There are different reasons why we consider them important in this study: 
\begin{itemize}
\item[\textit{A.}] Self-touch is likely to contribute to the formation of initial sensorimotor representations, and may therefore constitute one of the very first cues for subjective experience during early developmental stages. 

\item[\textit{B.}] The way the brain interprets action effects has been shown to differ depending on whether the sensory perception is self- or externally produced, with respect both to the perceived timing of their occurrence (intentional binding) and to their intensity (sensory attenuation) \cite{hughes2013mechanisms}.

\item[\textit{C.}] The rubber hand illusion has been extensively used as a paradigm for investigating the mechanisms underlying the sensorimotor minimal self.
\end{itemize}

An obvious advantage of conducting robotics research as test-beds for cognitive models in humans is that using robots allows unmediated access to the actual process (the algorithm) and information that is registered and processed in the system. In human research, even with advanced neuroimaging and analysis methods, researchers could only ever have an approximation of the actual cognitive processes underlying behaviour. This is especially true when investigating subjective experiences such as the illusion of a dummy hand or object being "mine", which is measured by observing behaviour or with subjective self-reports from human participants, whereas the difficulty lies with developing and most importantly validating objective measures of cognitive processes (e.g. neuroimaging, proprioceptive drift). The question of how we can measure a "subjective" experience in a robot then arises.


\subsection{Self-Touch}


\textit{How does self-touch relate to body ownership?}

Throughout development, the brain must establish links between sensory and motor maps. Refinement of these links eventually leads to goal directed actions. Establishing the basis of this lies in forming sensorimotor contingencies---the statistical links between sensory and motor information. 

These sensorimotor contingencies are thought to be established through body babbling where the infant moves its body in an exploratory manner, whereas the brain initiates actions and organises the resulting sensory outcomes continuously, refining its ability to predict sensory outcomes from motor actions. 
The process of establishing sensorimotor contingencies is gradual due to the large amount of information that is being processed---sensory inputs, motor outputs, and the statistical correlations between them.

Hoffmann et al. \cite{hoffmann2017development} suggest that the most systematic correlations are those that will emerge most easily, and therefore, the links between motor actions and proprioceptive changes are presumably the simplest to be extracted, followed by the links between motor actions and tactile input.
Hoffman \cite{hoffmann2017role} suggests that the redundant information arising from the configurations of self-touch in the proprioceptive-tactile space facilitates learning the body model in space. He asserts that pre-natal self-touch likely contributes to the formation of initial somatosensory representations. Evidence for an early integration between modalities comes from the instances of hand-to-mouth anticipation already in the womb \cite{myowa2006human,reissland2014development}. However, the formation of more comprehensive multimodal body representations probably occurs after birth, from 2-3 months, to include the visual modality and its connections to tactile-proprioceptive modalities. These are learned through self-exploration including self-touch within the environment, which involves learning temporal contingencies, spatial congruence, and redundancies of information coming from different sensory modalities. 

Rochat et al. \cite{rochat2000perceived} 
provide evidence that early on infants are capable of discriminating perceptual events---tactile stimuli---that are either self- or not self-produced. The authors tested the rooting behaviour---a reflex behaviour that is triggered by touching the cheek of the infant---in 24h and 4 weeks old newborns, and reported that infants tended to manifest rooting responses almost three times more often in response to external compared to self-stimulation.  This suggests that infants pick up already at birth sensorimotor contingencies (single touch or double touch) that specify self- versus external stimulation.

In a longitudinal study, Hofmann et al. \cite{hoffmann2017development} observed how infants between 3 and 21 months react to vibrotactile stimulation applied to different body parts. They report responses that varied between particular movement in the stimulated body part and successfully reaching and removing the buzzer. 
They found an overall developmental progression from general to specific movement patterns, especially from 3 to 12 months. Specifically, their results suggest that at 3-4 months, the infant responds to the buzzer in a non-specific way by moving its whole body, rather than moving the particular limb that was stimulated. However, between 4-12 months, the limb-specific buzzer-oriented reaching develops.

\textit{Robotics research}

Yamada et al. \cite{yamada2016embodied} worked on an embodied brain model of a human foetus in order to examine the causal link between sensorimotor experiences through embodied interactions and cortical learning of body representations. The embodied brain model was based on anatomical and physiological data and included a cortex, a spinal circuit, and musculoskeletal body with sensory receptors for proprioception, tactile perception, and vision, within a model of a uterine environment. The results of this study showed that embodied interaction between brain-body-environment within the womb help to guide the cortical learning of body representations through regularities in sensorimotor experiences. Also, the embodied interactions inside the womb provided a better arena for forming cortical body representations when compared to extrauterine embodied interactions. These findings support the notion suggested by previous studies on animal newborns and preterm infants that the formation of body representations begins even before birth. In addition, their findings suggest that embodied interactions inside the womb set the stage for visual-somatosensory integration after birth \cite{yamada2016embodied}. 

Expanding on this line of research but in a real robot body, Noda et al. \cite{noda2012super} equipped a humanoid robot with soft skin sensors and left it engaging in human-robot touch interactions. The tactile data generated through such interactions were used by the authors to form a self-organising somatosensory map, where the feature space was composed by spatially-adjacent sensor pairs. Computational body representations based on self-organising maps and multimodal integration through Hebbian connections have been proposed by Schillaci et al. \cite{lara2016body}, although not in the context of self-touch. Nonetheless, the model enabled a humanoid robot with predictive capabilities, fundamental for the study of sensory attenuation processes and sense of agency (see next section).

In \cite{hoffmann2018robotic} Hoffman et al. targeted the development of self-organising body representations in the iCub humanoid robot, learning somatosensory representations of the tactile space. 
Hoffmann et al. examined how the iCub robot, equipped with artificial skin, could form a topographic body representation by learning from tactile stimulations over the surface of its skin. They used modified SOMs that restrict the size of the maximum receptive field (MRF) of neuron groups at the output layer in order to reproduce the genetically predetermined organisation of somatotopic cortical maps. The formation of a tactile map organisation (predetermined with the MRF-SOM) has been reproduced by using training data obtained from a "double-touch" procedure, in which two experimenters provided tactile stimulation in two different places on the artificial skin.

\textit{Current limitations in robotics research}

The authors in \cite{hoffmann2018robotic} focused on mimicking the formation process of the cortical "homunculus" in a real robot, determined a-priori. The training data for the model was not generated by the robot itself, but rather with two methods: either with simulated data or with a human (or two, for studying multi-touch) experimenter touching the robot. Indeed, investigating self-touch in robotics would require the robot itself generating the 
behaviour that drives the formation of the body representation, which would incorporate the work on proprioceptive representations in \cite{hoffmann2016encoding}. Another possible expansion could be to implement this in a predictive coding framework, such that the self-touch behaviour would be driven by prediction error minimization. 



\subsection{Intentional binding and sensory attenuation} 

\textit{How do intentional binding and sensory attenuation relate to agency?}

The human brain interprets action effects differently depending on whether the sensory perception is self-produced or externally generated, with respect both to the perceived timing of their occurrence (intentional binding) and to their intensity (sensory attenuation) \cite{hughes2013mechanisms}.

In 2002, Haggard et al. studied the perceived time of intentional actions and their sensory consequences, while investigating action awareness \cite{haggard2002voluntary}. Interestingly, they found that voluntary actions and their effects are perceived as closer in time, compared to the perceived time shift between involuntary movements---induced by transcranial magnetic stimulation (TMS)---followed by the same effects. Specifically, subjects perceived voluntary movements as occurring later than when they actually occurred, and their sensory consequences as occurring earlier. This effect, known as \textit{intentional binding}, has attracted the attention of many scholars interested in shedding light on the nature of the sense of agency (see \cite{moore2012intentional} for a review).
Engel and Singer reported several pieces of evidence from animal and human studies suggesting that temporal dynamics in neuronal activity may be critically involved in conscious states, in particular that synchronisation may be involved in the generation and maintenance of sensory awareness \cite{engel2001temporal}.

\textit{Sensory attenuation} refers to the partial cancellation, or reduction, of the perceived intensity of the effects of a self-initiated action. Several studies show similar effects.
Blakemore et al., for instance, found that self-produced tactile stimulation was perceived as less intense compared to when the same stimulus was produced externally \cite{blakemore1999spatio}. 
In the experiment, participants moved the arm of a robot with their left hand in order to produce tactile stimuli on their right hand via a second robot. The authors found that varying the delay between the movements of the left hand and the resulting movements producing the tactile stimuli on the right hand, and varying the degree of trajectory perturbation all had an effect on the rating of tickliness sensations. Participants perceived the stimuli produced by more delayed and more perturbated movements as more tickling, suggesting that self-produced movements attenuate resulting sensations and that a prerequisite for this attenuation is that stimuli and their causal motor commands correspond in time and space.


\textit{Robotics research}

In \cite{michel2004motion}, Michel et al. experimented the incremental learning of characteristic time delay inherent in the action-perception loop from a sequence of random arm motions within the visual field in a humanoid robot. Interestingly, the study showed that the learned time delay can be successfully used to identify own body parts in the visual field. 

Lang, Schillaci and Hafner \cite{Lang2018} studied how a humanoid robot can learn, through a self-exploration behaviour, the sensory outcomes (in the visual domain) of self-generated movements. The  sensorimotor experience gathered during this process was used as training data for a deep convolutional neural network that mapped proprioceptive and motor data (e.g. initial arm joint positions and applied motor commands) onto the visual outcomes of these actions. The authors then used such a forward model in two experiments. First, for generating visual predictions of self-generated movements, which were compared to actual visual perceptions and then used to compute a prediction error. The system generated higher prediction errors when an external subject was performing actions in front of the robot, compared to situations in which the robot was observing only itself doing the same arm movements. 
The authors also showed how predictions can be used to attenuate self-generated movements, and thus create enhanced visual perceptions, where the sight of objects---originally occluded by the robot body---was still maintained. This suggests that similar processes may shed light also on the understanding of the sense of object permanence and of short term memory systems in humans.

In \cite{lara2016body}, Schillaci et al. presented a biologically inspired model for learning multimodal body representations in artificial agents in the context of learning and predicting robot ego-noise, i.e. the auditory noise produced by the robot's motors while it moves. The authors performed an ego-noise attenuation experiment, which showed the effects in the ego-noise suppression performance of coherent and incoherent proprioceptive and motor information passed as inputs to the predictive process implemented by a forward model. In line with the aforementioned studies, the experiments showed that ego-noise attenuation was more pronounced when the robot was the \textit{owner} of the action. Sensory attenuation was less pronounced when the robot (\textit{not moving} itself) was listening to a simulated moving robot, as the incongruence of the proprioceptive and motor information with the perceived ego-noise generated larger prediction errors. 

Other robotics studies can be found in the literature implementing top-down processes for \textit{interpreting} bottom-up sensory streams. An example is the interesting work of Jun Tani \cite{tani1998interpretation}, who implemented an incremental learning system based on recurrent neural networks and self-organising networks that evolves by showing steady and unsteady phases. The author explains these fluctuations as a result of the interaction between top-down and bottom-up processes, and makes a parallel between them and phenomenological observations.

\textit{Current limitations in robotics research}

For many years, sense of agency has been measured using explicit self-reported judgements (e.g. \cite{daprati2007kinematic}). However, the self-report approach has limitations, as it is sensitive to different biases \cite{cavazzana2016sense}. 
Moreover, this is not a feasible approach in current robotics technologies if we look at it from a phenomenological perspective, as the only level of judgement available in robots comes at the point when the experimenter observes and interprets the internal states of the machine.

Tani \cite{tani1998interpretation}, for instance, compares the dynamical structure of the system created in his robotic study to the structure of the ''self'', and observes that ''the ''self'' is made aware when the unsteady phase appears in the course of the time-development of the system". Although we recognise the quality of the proposed model, we believe that similar statements on subjective mind and self in robots---as reported in the paper---are prone to criticisms about their objective validity. 

In the work of Michel et al. on learning time delays in action-perception loops in humanoid robots \cite{michel2004motion}, the learning mechanism was simply looking for regularities in the delays between the motor activations and the detections of moving objects on the screen. The authors performed a basic processing of the visual input, which was at any stage mapped to other sensory modalities (e.g. to proprioceptive modalities). The authors pointed out the need for a more robust method of visual recognition of the robot body. However, we find as more critical the lack of multimodal integration, which as to the proposed study consisted only in the analysis of the timestamps of intermodal events. In fact, as discussed in section \ref{sec:development}, the development of multimodal body representations is a necessary condition for the emergence of a minimal self. Further works on intentional binding effects in the investigation of sense of body ownership and of agency in robots should thus consider also this aspect.

In general, there is a need for further investigation on the intentional binding action effects, especially in robotics. In fact, there are studies \cite{desantis2012intentional} showing that motor prediction seems to not modulate that, casting doubts on the assumption that intentional binding of action effects is linked to an internal forward predictive process. These studies suggest that just the temporal control of a stimulus, by means of a voluntary action, might be sufficient to trigger the binding effect.

Regarding the sensory attenuation measure, most of the robotics studies addressing this topic (e.g. \cite{lara2016body, Lang2018}) adopt comparator computation models for sensory prediction and attenuation. Nonetheless, more recent probabilistic computational proposals---such as the predictive coding framework described in the next section---would represent a more biologically plausible approach. 

\subsection{Rubber Hand Illusion (RHI)}

\textit{How does the RHI relate to body ownership?}

The rubber hand illusion (RHI) \cite{botvinick1998rubber}, along with other body ownership illusions, is a widely used paradigm for investigating the mechanisms underlying the (sensorimotor) minimal self. In the rubber hand illusion, an observer sees a dummy hand receiving touch (e.g. brush strokes) while at the same time receiving tactile stimulation on their real hidden hand at the same location. This usually elicits an illusory experience of sensing the stimulation on the dummy hand \cite{botvinick1998rubber, ehrsson2004s}, and as a result, incorporating the fake hand as a part of the observer's own body \cite{tsakiris2005rubber, tsakiris2010my}. This effect was found to be reflected behaviourally as a fear response when the fake hand is being threatened \cite{armel2003projecting, ehrsson2007threatening} or in perceiving the location of the real hand as closer towards where the dummy hand is located (a "proprioceptive drift") \cite{botvinick1998rubber, tsakiris2005rubber}, suggesting that the dummy hand is treated by the brain as part of the own body as a result of the multisensory stimulation \cite{tsakiris2010my}. These illusory effects are thought to be an indicator for the presence of multimodal, integrated body representations.
RHI studies show that both top-down and bottom-up processes influence the embodiment of the dummy hand. For example, top-down expectations about the appearance of the human hand---stronger illusory effects are experienced when the dummy hand closely resembles a human hand---are thought to result from internal body representations, and bottom-up sensory inputs where illusory effects are dependent on spatiotemporal congruence of the stimulation and on the proximity of the dummy hand to the real hand \cite{tsakiris2005rubber}.

\textit{A predictive coding account of the RHI} 

Different explanations to the RHI effects have been proposed in the cognitive neuroscience literature.
The \textit{predictive coding} \cite{friston2005theory} account, that is recently receiving great support, proposes that this results from the fact that, to reduce uncertainty, the brain makes inferences about causes of sensory events in a probabilistic-Bayesian manner: prior beliefs (bias) represented in internal models generate predictions about sensory input (top-down). When predictions contradict actual sensory input, this generates "prediction errors" that propagate up the hierarchy---to unimodal, multimodal, representational areas (bottom-up). The contradiction results in "surprise" that needs to be "explained away" by updating the model, thus reducing the prediction error (see \cite{limanowski2013minimal} for a review on the minimal self within the predictive coding framework). 

During the RHI, the co-occurrence of the visual input that comes from observing touch on the dummy hand together with the tactile input that comes from the stimulation of the real hand, evokes---according to the predictive coding account---a prediction error (or surprise), because this spatiotemporal congruence is not predicted by the initial generative forward model. According to \cite{apps2014free}, the illusion is induced when the probability of the dummy hand being "me" exceeds the probability of the real hand being "one's own", given the sensory input. If the prediction error can be explained away by adjusting the body model to incorporate the dummy hand, then the RHI will be induced. The explanation that the dummy hand is "mine" is equivalent to mean that the visual and tactile perceptual inputs occur at the same location and arise from a common cause. 


\textit{RHI and sense of agency: "passive" versus "active" RHI}

In the classic RHI paradigm, the participants are not allowed to move their hand. It has been observed that if they do move their hand, and they do not observe congruent movement in the dummy hand, then the illusion will be immediately abolished. The proposed explanation is that, as the participant moved their hand in order to "test" the body ownership of the \textit{not moving} dummy hand, the prediction error cannot resolve in favour of perceiving the dummy hand as one's own \cite{limanowski2013minimal}. This classic RHI paradigm is therefore named as "passive", and while it induces the illusion of body ownership, the effect it has on the sense of agency can not be directly examined.
 
In "active" RHI, in addition to the concurrent multisensory stimulation, the participant also moves their real hand while observing a dummy hand that moves along it \cite{slater2009inducing, kalckert2012moving}. In this version of the RHI paradigm (which could also be induced in a virtual environment), the illusion of embodying a dummy hand (or object) is induced as a result of the congruence between the participant's motor actions and the sensory outcomes of said actions, namely, the perception of the movement of the dummy hand, rather than as a result of the multimodal sensory integration alone. Also of note, in this case, the visual properties of the dummy hand or object, do not necessarily have to resemble those of a real human hand \cite{slater2009inducing, ma2015role}. In "active" RHI, one can directly manipulate the sense of agency, or even possibly disassociate agency from body ownership \cite{kalckert2012moving}. In line with this, there is evidence that body ownership or embodiment of an object, even one which is anatomically implausible, can still be successfully induced given systematic synchrony between visual input when observing the object and one's own movement. In \cite{ma2015body}, Ma and Hommel induced in a virtual setting body ownership of virtual 2-D shapes (a virtual balloon changing in size, and a virtual square changing in size and color) when the changes in the 2-D shapes were systematically congruent with participants' actions. In addition to the concurrent multisensory stimulation, the induced illusion of body ownership is thought to be cultivated by the congruence between predicted sensory outcomes of motor actions and the actual sensory input, pointing to the role of agency in body ownership. This is reminiscent of the manner in which the sense of body ownership emerges in the ontogenetic developmental process.

In another study in a virtual setting \cite{ma2015role}, Ma and Hommel manipulated the similarity of the object-to-be-embodied (end-effector) to the real hand, the synchrony between stimulation or movement of the end-effector and the stimulation or movement of participant's real hand, and the degree of agency, operationalized by the level of control over the end-effector. They found that agency strongly effected synchrony-induced body ownership, but not similarity. However, both similarity and agency induced a bias towards body ownership of the end-effector. This shows that agency contributes to body ownership.


\textit{Robotics research}

When humans are subjected to the RHI, they show a perceptual drift in the location of the real hand toward the dummy hand, which suggests an update in the body representation. Using a multisensory robotic arm, Hinz et al. \cite{hinz2018drifting} replicated these drifting patterns in both human and robot experiments with the classic ("passive") RHI paradigm. The learning and estimation algorithm \cite{lanillos2018adaptive} used in the study was based on the framework of predictive coding \cite{friston2005theory}. Specifically, Lanillos and Cheng \cite{lanillos2018adaptive} developed a method for integrating different sources of information (tactile, visual, and proprioceptive) that drives the robot priors to infer its body configuration. This computational perceptual model enables a multisensory robot to learn, make inferences, and update its body configuration from its sensors. They modeled the robot body estimation as a process of minimizing the prediction error between the body configuration "belief" (prediction) and the observed posterior, and minimizing the variational free energy \cite{friston2005theory} by using the sensory prediction error. 
Using the algorithm in \cite{lanillos2018adaptive}, Hinz et al. \cite{hinz2018drifting} showed that body configuration estimation can be done through minimization of prediction error as one process that involves both predictive coding and causal inference. The results from the human and robot experiments suggest that the perceived locations of both the real and the dummy hand drift to a common location between them. In human data, in fact, illusion scores (self-report) were not correlated with the proprioceptive drift, suggesting that the drift and body-ownership illusions are related, but different processes \cite{abdulkarim2016no}.

\textit{Current limitations in robotics research}

Many studies can be found in the literature which use robotic or virtual hands in active RHI experiments, that is in scenarios where participants move their real hand while observing a dummy robotic or virtual hand that moves along it. However, the investigation on RHI \textit{experienced by artificial systems} is very scarce. To the best of our knowledge, the work of Hinz et al. \cite{hinz2018drifting} is the only study on replicating the rubber hand illusion on a robot. 

Another concern is related to the "proprioceptive drift" as the classic "objective measure" for the RHI, as used in the experiment mentioned above \cite{hinz2018drifting}. Both the human data from this study, as well as previous work in humans \cite{abdulkarim2016no, holle2011proprioceptive}, failed to find a correlation between the proprioceptive drift and the self-report of the participants, casting doubt on the validity of the proprioceptive drift as an objective measure for body ownership. Further investigation is therefore suggested.
Also, after reproducing the classic, "passive" RHI in a robot using free energy minimization \cite{hinz2018drifting}, it could be suggested to examine the algorithm in an "active" RHI experimental setup, which would allow to distinctly examine sense of agency in a robot apart from body ownership. 

\section{Conclusions}
\label{sec:discussion}
This manuscript presented an interdisciplinary overview of developmental indices and behavioural measures of the minimal self. The fundamental role of the development of body representation in the emergence of body ownership and sense of agency has been discussed.
This work also addressed the task of experimentally quantifying the attribution of subjective experience, and surveyed a number of behavioural paradigms and measures indicating the presence of different aspects of the minimal self, namely self-touch, intentional binding and sensory attenuation, and the rubber hand illusion. 

Self-touch is likely to contribute to the formation of initial sensorimotor representations, and may therefore constitute one of the very first cues for subjective experience during early developmental stages. Moreover, the way in which our brain interprets action effects has been shown to differ depending on whether the sensory perception is self-produced or externally triggered, with respect both to the perceived timing of their occurrence (intentional binding) and to their intensity (sensory attenuation). Finally, we addressed the rubber hand illusion as it has been extensively used as a paradigm for investigating the mechanisms underlying the sensorimotor minimal self.

We reviewed the most prominent studies addressing these paradigms and measures from the literature in neuroscience, cognitive and developmental sciences. For each of these topics, we presented related robotics studies. Equipping robots with self-awareness and studying the possibility of subjective experience in artificial systems is, in fact, of high interest for the cognitive and developmental robotics communities. This manuscript contributed to this quest by identifying current knowledge gaps and limitations in robotics. In the next section, we conclude this work by highlighting the most critical gaps and by suggesting further research directions.

\textit{Further research directions in robotics}

The development of multimodal body representations has been discussed as fundamental in the emergence of self-awareness. Further robotics research should address the implementation of multimodal integration through online developmental processes.
Current research on self-touch and on self-organisation of somatosensory maps in robots do not explicitly consider the \textit{active} role that the robot should have in the generation of sensorimotor experience. We therefore encourage further experimentation considering self-exploration behaviours in such a developmental process.

Recent proposals on predictive processes represent promising research lines that go beyond their higher level of biological plausibility. Prediction error minimization processes could result in intelligent robot exploration behaviours, where the intrinsic motivation of reducing uncertainty would generate artificial curiosity and goal-directed behaviours---both prerequisites for motor and cognitive development.

The intentional binding effects and sensory attenuation processes are recognised by the neuroscience and cognitive science communities as important measures for the definition of self-boundaries. 
Current studies, however, mostly focus on explicit judgement from human participants. This self-report approach is clearly not feasible in robotics. Nonetheless, robots allow experimenters to inspect their internal states, the flowing sensorimotor data and the predictive processes implemented by their computational models. These data is for obvious reasons not accessible in humans. Robots represent, therefore, promising tools for the investigation of intentional binding effects and sensory attenuation processes. Beside encouraging further investigations in robotics, we also suggest more experimentations considering the effects of the developmental path in the performance of such measures. In particular, how do developmental stages---for instance, the levels of multimodal integration reached after a certain stage of sensorimotor exploration---affect predictive performances, and consequently sensory attenuation and intentional binding effects? Can this be linked to stages in early development of the minimal self in humans?

The RHI represents a well-established paradigm to measure subjective experience in humans, and it makes therefore very much sense to extend this to the investigation of subjective experiences in robots. Further usage of this measure in the study of the artificial self is thus encouraged. In particular, we suggest testing the "active" RHI paradigm in robots in order to investigate also the sense of agency. Moreover, similar effects as the ones mentioned above could be studied for the RHI paradigm. Further studies could address whether and how the perceptual drift measure is affected by the developmental stage in which the agent finds themselves.






\bibliographystyle{IEEEtran} 
\bibliography{biblio}

\begin{thebibliography}{10}
\providecommand{\url}[1]{#1}
\csname url@samestyle\endcsname
\providecommand{\newblock}{\relax}
\providecommand{\bibinfo}[2]{#2}
\providecommand{\BIBentrySTDinterwordspacing}{\spaceskip=0pt\relax}
\providecommand{\BIBentryALTinterwordstretchfactor}{4}
\providecommand{\BIBentryALTinterwordspacing}{\spaceskip=\fontdimen2\font plus
\BIBentryALTinterwordstretchfactor\fontdimen3\font minus
  \fontdimen4\font\relax}
\providecommand{\BIBforeignlanguage}[2]{{%
\expandafter\ifx\csname l@#1\endcsname\relax
\typeout{** WARNING: IEEEtran.bst: No hyphenation pattern has been}%
\typeout{** loaded for the language `#1'. Using the pattern for}%
\typeout{** the default language instead.}%
\else
\language=\csname l@#1\endcsname
\fi
#2}}
\providecommand{\BIBdecl}{\relax}
\BIBdecl

\bibitem{gallagher2000philosophical}
S.~Gallagher, ``Philosophical conceptions of the self: implications for
  cognitive science,'' \emph{Trends in cognitive sciences}, vol.~4, no.~1, pp.
  14--21, 2000.

\bibitem{martin2014temporal}
B.~Martin, M.~Wittmann, N.~Franck, M.~Cermolacce, F.~Berna, and A.~Giersch,
  ``Temporal structure of consciousness and minimal self in schizophrenia,''
  \emph{Frontiers in psychology}, vol.~5, p. 1175, 2014.

\bibitem{ferri2011motor}
F.~Ferri, F.~Frassinetti, M.~Costantini, and V.~Gallese, ``Motor simulation and
  the bodily self,'' \emph{PloS one}, vol.~6, no.~3, p. e17927, 2011.

\bibitem{damasio2003mental}
A.~Damasio, ``Mental self: The person within,'' \emph{Nature}, vol. 423, no.
  6937, p. 227, 2003.

\bibitem{engel2001temporal}
A.~K. Engel and W.~Singer, ``Temporal binding and the neural correlates of
  sensory awareness,'' \emph{Trends in cognitive sciences}, vol.~5, no.~1, pp.
  16--25, 2001.

\bibitem{crick1990towards}
F.~Crick and C.~Koch, ``Towards a neurobiological theory of consciousness,'' in
  \emph{Seminars in the Neurosciences}, vol.~2.\hskip 1em plus 0.5em minus
  0.4em\relax Saunders Scientific Publications, 1990, pp. 263--275.

\bibitem{rochat2003five}
P.~Rochat, ``Five levels of self-awareness as they unfold early in life,''
  \emph{Consciousness and cognition}, vol.~12, no.~4, pp. 717--731, 2003.

\bibitem{sep-bodily-awareness}
F.~de~Vignemont, ``Bodily awareness,'' in \emph{The Stanford Encyclopedia of
  Philosophy}, spring 2018~ed., E.~N. Zalta, Ed.\hskip 1em plus 0.5em minus
  0.4em\relax Metaphysics Research Lab, Stanford University, 2018.

\bibitem{gallagher1986body}
S.~Gallagher, ``Body image and body schema: A conceptual clarification,''
  \emph{The Journal of Mind and Behavior}, pp. 541--554, 1986.

\bibitem{dall2018somatotopic}
S.~Dall’Orso, J.~Steinweg, A.~Allievi, A.~Edwards, E.~Burdet, and T.~Arichi,
  ``Somatotopic mapping of the developing sensorimotor cortex in the preterm
  human brain,'' \emph{Cerebral Cortex}, vol.~28, no.~7, pp. 2507--2515, 2018.

\bibitem{bradley1975fetal}
R.~M. Bradley and C.~M. Mistretta, ``Fetal sensory receptors,''
  \emph{Physiological Reviews}, vol.~55, no.~3, pp. 352--382, 1975.

\bibitem{fagard2018fetal}
J.~Fagard, R.~Esseily, L.~Jacquey, K.~O’Regan, and E.~Somogyi, ``Fetal origin
  of sensorimotor behavior,'' \emph{Frontiers in neurorobotics}, vol.~12, 2018.

\bibitem{myowa2006human}
M.~Myowa-Yamakoshi and H.~Takeshita, ``Do human fetuses anticipate
  self-oriented actions? a study by four-dimensional (4d) ultrasonography,''
  \emph{Infancy}, vol.~10, no.~3, pp. 289--301, 2006.

\bibitem{reissland2014development}
N.~Reissland, B.~Francis, E.~Aydin, J.~Mason, and B.~Schaal, ``The development
  of anticipation in the fetus: a longitudinal account of human fetal mouth
  movements in reaction to and anticipation of touch,'' \emph{Developmental
  psychobiology}, vol.~56, no.~5, pp. 955--963, 2014.

\bibitem{zoia2007evidence}
S.~Zoia, L.~Blason, G.~D’Ottavio, M.~Bulgheroni, E.~Pezzetta, A.~Scabar, and
  U.~Castiello, ``Evidence of early development of action planning in the human
  foetus: a kinematic study,'' \emph{Experimental Brain Research}, vol. 176,
  no.~2, pp. 217--226, 2007.

\bibitem{hoffmann2017role}
M.~Hoffmann, ``The role of self-touch experience in the formation of the
  self,'' \emph{arXiv preprint arXiv:1712.07843}, 2017.

\bibitem{mcgraw1943neuromuscular}
M.~B. McGraw, ``The neuromuscular maturation of the human infant.'' \emph{Yale
  Journal of Biology and Medicine}, vol.~15, no.~6, 1943.

\bibitem{chugani1994development}
H.~T. Chugani, ``Development of regional brain glucose metabolism in relation
  to behavior and plasticity.'' \emph{Human behavior and the developing brain},
  pp. 153--175, 1994.

\bibitem{thomas2015independent}
B.~L. Thomas, J.~M. Karl, and I.~Q. Whishaw, ``Independent development of the
  reach and the grasp in spontaneous self-touching by human infants in the
  first 6 months,'' \emph{Frontiers in psychology}, vol.~5, p. 1526, 2015.

\bibitem{blakemore1999spatio}
S.-J. Blakemore, C.~D. Frith, and D.~M. Wolpert, ``Spatio-temporal prediction
  modulates the perception of self-produced stimuli,'' \emph{Journal of
  cognitive neuroscience}, vol.~11, no.~5, pp. 551--559, 1999.

\bibitem{friston2005theory}
K.~Friston, ``A theory of cortical responses,'' \emph{Phil. trans. of the Royal
  Society B: Biological sciences}, vol. 360, no. 1456, pp. 815--836, 2005.

\bibitem{rochat1998self}
P.~Rochat, ``Self-perception and action in infancy,'' \emph{Experimental brain
  research}, vol. 123, no. 1-2, pp. 102--109, 1998.

\bibitem{hughes2013mechanisms}
G.~Hughes, A.~Desantis, and F.~Waszak, ``Mechanisms of intentional binding and
  sensory attenuation: The role of temporal prediction, temporal control,
  identity prediction, and motor prediction.'' \emph{Psychological bulletin},
  vol. 139, no.~1, p. 133, 2013.

\bibitem{hoffmann2017development}
M.~Hoffmann, L.~K. Chinn, E.~Somogyi, T.~Heed, J.~Fagard, J.~J. Lockman, and
  J.~K. O’Regan, ``Development of reaching to the body in early infancy: From
  experiments to robotic models,'' in \emph{IEEE Int. Conf. on Development and
  Learning and Epigenetic Robotics}.\hskip 1em plus 0.5em minus 0.4em\relax
  IEEE, 2017.

\bibitem{rochat2000perceived}
P.~Rochat and T.~Striano, ``Perceived self in infancy,'' \emph{Infant behavior
  and development}, vol.~23, no. 3-4, pp. 513--530, 2000.

\bibitem{yamada2016embodied}
Y.~Yamada, H.~Kanazawa, S.~Iwasaki, Y.~Tsukahara, O.~Iwata, S.~Yamada, and
  Y.~Kuniyoshi, ``An embodied brain model of the human foetus,''
  \emph{Scientific reports}, vol.~6, p. 27893, 2016.

\bibitem{noda2012super}
T.~Noda, T.~Miyashita, H.~Ishiguro, and N.~Hagita, ``Super-flexible skin
  sensors embedded on the whole body, self-organizing based on haptic
  interactions,'' \emph{Human-Robot Interaction in Social Robotics}, p. 183,
  2012.

\bibitem{lara2016body}
B.~Lara, V.~V. Hafner, C.-N. Ritter, and G.~Schillaci, ``Body representations
  for robot ego-noise modelling and prediction. towards the development of a
  sense of agency in artificial agents,'' in \emph{Proc. of the Artificial Life
  Conference 2016 13}.\hskip 1em plus 0.5em minus 0.4em\relax MIT Press, 2016,
  pp. 390--397.

\bibitem{hoffmann2018robotic}
M.~Hoffmann, Z.~Straka, I.~Farka{\v{s}}, M.~Vavre{\v{c}}ka, and G.~Metta,
  ``Robotic homunculus: Learning of artificial skin representation in a
  humanoid robot motivated by primary somatosensory cortex,'' \emph{IEEE Trans.
  on Cognitive and Developmental Systems}, vol.~10, no.~2, pp. 163--176, 2018.

\bibitem{hoffmann2016encoding}
M.~Hoffmann and N.~Bednarova, ``The encoding of proprioceptive inputs in the
  brain: knowns and unknowns from a robotic perspective,'' \emph{arXiv preprint
  arXiv:1607.05944}, 2016.

\bibitem{haggard2002voluntary}
P.~Haggard, S.~Clark, and J.~Kalogeras, ``Voluntary action and conscious
  awareness,'' \emph{Nature neuroscience}, vol.~5, no.~4, p. 382, 2002.

\bibitem{moore2012intentional}
J.~W. Moore and S.~S. Obhi, ``Intentional binding and the sense of agency: a
  review,'' \emph{Consciousness and cognition}, vol.~21, no.~1, pp. 546--561,
  2012.

\bibitem{michel2004motion}
P.~Michel, K.~Gold, and B.~Scassellati, ``Motion-based robotic
  self-recognition,'' in \emph{IEEE/RSJ International Conference on Intelligent
  Robots and Systems}, vol.~3.\hskip 1em plus 0.5em minus 0.4em\relax IEEE,
  2004, pp. 2763--2768.

\bibitem{Lang2018}
C.~{Lang}, G.~{Schillaci}, and V.~V. {Hafner}, ``A deep convolutional neural
  network model for sense of agency and object permanence in robots,'' in
  \emph{Joint IEEE International Conference on Development and Learning and
  Epigenetic Robotics}, 2018.

\bibitem{tani1998interpretation}
J.~Tani, ``An interpretation of the ‘self’from the dynamical systems
  perspective: A constructivist approach,'' \emph{Journal of Consciousness
  Studies}, vol.~5, no. 5-6, pp. 516--542, 1998.

\bibitem{daprati2007kinematic}
E.~Daprati, S.~Wriessnegger, and F.~Lacquaniti, ``Kinematic cues and
  recognition of self-generated actions,'' \emph{Experimental Brain Research},
  vol. 177, no.~1, pp. 31--44, 2007.

\bibitem{cavazzana2016sense}
A.~Cavazzana, ``Sense of agency and intentional binding: How does the brain
  link voluntary actions with their consequences?'' \emph{Ph.D. thesis,
  Dipartimento di Psicologia Generale, University of Padua, Italy}, 2016.

\bibitem{desantis2012intentional}
A.~Desantis, G.~Hughes, and F.~Waszak, ``Intentional binding is driven by the
  mere presence of an action and not by motor prediction,'' \emph{PLoS one},
  vol.~7, no.~1, p. e29557, 2012.

\bibitem{botvinick1998rubber}
M.~Botvinick and J.~Cohen, ``Rubber hands ‘feel’touch that eyes see,''
  \emph{Nature}, vol. 391, no. 6669, p. 756, 1998.

\bibitem{ehrsson2004s}
H.~H. Ehrsson, C.~Spence, and R.~E. Passingham, ``That's my hand! activity in
  premotor cortex reflects feeling of ownership of a limb,'' \emph{Science},
  vol. 305, no. 5685, pp. 875--877, 2004.

\bibitem{tsakiris2005rubber}
M.~Tsakiris and P.~Haggard, ``The rubber hand illusion revisited: visuotactile
  integration and self-attribution.'' \emph{Journal of Exp. Psychology: Human
  Perception and Performance}, vol.~31, no.~1, p.~80, 2005.

\bibitem{tsakiris2010my}
M.~Tsakiris, ``My body in the brain: a neurocognitive model of
  body-ownership,'' \emph{Neuropsychologia}, vol.~48, no.~3, pp. 703--712,
  2010.

\bibitem{armel2003projecting}
K.~C. Armel and V.~S. Ramachandran, ``Projecting sensations to external
  objects: evidence from skin conductance response,'' \emph{Proc. of the Royal
  Society of London. Series B: Biological Sciences}, vol. 270, no. 1523, pp.
  1499--1506, 2003.

\bibitem{ehrsson2007threatening}
H.~H. Ehrsson, K.~Wiech, N.~Weiskopf, R.~J. Dolan, and R.~E. Passingham,
  ``Threatening a rubber hand that you feel is yours elicits a cortical anxiety
  response,'' \emph{Proc. of the National Academy of Sciences}, vol. 104,
  no.~23, pp. 9828--9833, 2007.

\bibitem{limanowski2013minimal}
J.~Limanowski and F.~Blankenburg, ``Minimal self-models and the free energy
  principle,'' \emph{Frontiers in human neuroscience}, vol.~7, p. 547, 2013.

\bibitem{apps2014free}
M.~A. Apps and M.~Tsakiris, ``The free-energy self: a predictive coding account
  of self-recognition,'' \emph{Neuroscience \& Biobehavioral Reviews}, vol.~41,
  pp. 85--97, 2014.

\bibitem{slater2009inducing}
M.~Slater, D.~P{\'e}rez~Marcos, H.~Ehrsson, and M.~V. Sanchez-Vives, ``Inducing
  illusory ownership of a virtual body,'' \emph{Frontiers in neuroscience},
  vol.~3, p.~29, 2009.

\bibitem{kalckert2012moving}
A.~Kalckert and H.~H. Ehrsson, ``Moving a rubber hand that feels like your own:
  a dissociation of ownership and agency,'' \emph{Frontiers in human
  neuroscience}, vol.~6, p.~40, 2012.

\bibitem{ma2015role}
K.~Ma and B.~Hommel, ``The role of agency for perceived ownership in the
  virtual hand illusion,'' \emph{Consciousness and cognition}, vol.~36, pp.
  277--288, 2015.

\bibitem{ma2015body}
------, ``Body-ownership for actively operated non-corporeal objects,''
  \emph{Consciousness and cognition}, vol.~36, pp. 75--86, 2015.

\bibitem{hinz2018drifting}
N.-A. Hinz, P.~Lanillos, H.~Mueller, and G.~Cheng, ``Drifting perceptual
  patterns suggest prediction errors fusion rather than hypothesis selection:
  replicating the rubber-hand illusion on a robot,'' \emph{arXiv preprint
  arXiv:1806.06809}, 2018.

\bibitem{lanillos2018adaptive}
P.~Lanillos and G.~Cheng, ``Adaptive robot body learning and estimation through
  predictive coding,'' \emph{arXiv preprint arXiv:1805.03104}, 2018.

\bibitem{abdulkarim2016no}
Z.~Abdulkarim and H.~H. Ehrsson, ``No causal link between changes in hand
  position sense and feeling of limb ownership in the rubber hand illusion,''
  \emph{Attention, Perception, \& Psychophysics}, vol.~78, no.~2, pp. 707--720,
  2016.

\bibitem{holle2011proprioceptive}
H.~Holle, N.~McLatchie, S.~Maurer, and J.~Ward, ``Proprioceptive drift without
  illusions of ownership for rotated hands in the “rubber hand illusion”
  paradigm,'' \emph{Cognitive neuroscience}, vol.~2, no. 3-4, pp. 171--178,
  2011.

\end{thebibliography}

\end{document}